\PassOptionsToPackage{numbers,sort&compress}{natbib}
\documentclass[final,12pt,numbers]{elsarticle}

\usepackage{amssymb}

\usepackage{graphicx}
\usepackage{hyperref}
\usepackage[labelfont=bf]{caption}  
\usepackage{float}
\usepackage{multirow}
\usepackage[margin=1in]{geometry}
\usepackage{amsmath}

\begin{document}
\begin{frontmatter}

\title{
Physics-informed Kolmogorov–Arnold networks (PIKANs) for noise-robust reconstruction of boiling heat flux from infrared measurements
}

\author[a]{Siavash Khodakarami}
\author[b]{Parimala Vardhan Vydyula}
\author[a]{Aniruddha Bora}
\author [b]{Justin A. Weibel}
\author[a]{George Em Karniadakis\corref{cor1}}


\affiliation[a]{organization={Division of Applied Mathematics, Brown University},  
                city={Providence},  
                state={RI}, 
                postcode={02912},
                country={USA}}

\affiliation[b]{organization={School of Mechanical Engineering, Purdue University}, 
                city={West Lafayette}, 
                state={IN}, 
                postcode={47907},
                country={USA}}

\cortext[cor1]{Corresponding author. Email: george\_karniadakis@brown.edu}

\begin{abstract}
Recovering a noise-amplified inverse quantity from corrupted boundary data is a recurring problem in physics-informed learning for real-world applications. Specifically,  small perturbations in the measured field are magnified through the spatial derivatives needed to reconstruct the target, and hence conventional discretization solvers turn them into large unphysical oscillations. We study this problem through the reconstruction of local time-resolved nucleate-boiling heat flux, which cannot be measured directly and must be inferred from infrared surface temperatures by solving an unsteady three-dimensional heat-conduction problem with sharp and rapidly evolving liquid–vapor interface gradients. We show that neither network architecture nor optimizer alone overcomes the spectral bias that smooths away these interface-scale features, and that the two must be addressed jointly. Our approach combines a Chebyshev physics-informed Kolmogorov–Arnold network (PIKAN) backbone, a quasi-second-order (SOAP) optimizer, and a temperature-annealed residual-based resampling scheme that concentrates collocation points at high-residual interface regions. Trained directly on experimental infrared measurements imposed as a boundary condition, the network resolves high-frequency bubble signatures and the high-flux distribution tail much better than a Fourier-feature PINN of equal size and training budget. In particular, it selectively filters measurement noise while preserving the physically meaningful high-frequency features, suppressing the spurious oscillations that contaminate finite-volume reconstructions and yielding more physically consistent heat-flux estimates from noisy experimental data.

\end{abstract}
\end{frontmatter}

\section{Introduction}

{
\noindent Physics-informed neural networks (PINNs) have emerged as powerful class of scientific machine learning tools that combine data-driven deep learning with governing physical laws. PINNs embed partial differential equations (PDEs), boundary conditions, and initial conditions directly into the training objective of neural networks. Since their introduction by Raissi et al. \cite{raissi2019physics}, PINNs have been widely used for solving forward or inverse problems in variety of fields such as fluid mechanics, solid mechanics, and biomechanics. \cite{wang2025simulating}\\
}

{
Although PINNs have demonstrated considerable success on canonical scientific and engineering problems, their application to real-world engineering systems and high-dimensional problems remains challenging. Many PDE-governed phenomena involve widely separated spatial and temporal scales, high-frequency features, sharp gradients, or multiscale non-linear interactions. Examples of such problems are commonly observed in turbulence, shock formation, multiphase interfaces, and oscillating wave fields. Classical neural networks, generally defined as multi-layer perceptrons (MLPs), are known to have spectral bias toward low wavenumbers. \cite{rahaman2019spectral} As a result, they tend to learn low-frequency or smooth components of the target solution early in the training, while high-frequency components are learned significantly later, if at all. While spectral bias of MLPs has been widely recognized, the same problem exist for other neural networks including convolutional neural networks (CNNs) and transformers. \cite{khodakarami2025mitigating,wang2022anti} PINNs typically rely on MLP backbones and consequently exhibit spectral bias, and tend to converge prematurely to over-smoothed approximations of the solution, even when equipped with adequate training data and sufficient collocation points. This phenomenon can explain why training PINNs become challenging in stiff problems or chaotic systems.\\
}

{Another difficulty in training PINNs lies in the interaction between the loss formulation and the spectral properties of the neural networks. The PINN objective typically consist of multiple terms including initial and boundary conditions, governing PDEs, and (if available) data loss. Successful optimization of PINNs requires careful selection of weighting strategy for these terms. The neural network's inherent spectral bias further exacerbates this issue, yielding a highly non-convex optimization landscape that may encourage PINNs to converge to trivial and overly smoothed solutions. Researchers have proposed adaptive global and local weighting methods to facilitate optimization. Anagnostopoulos et al. \cite{anagnostopoulos2024residual} proposed residual-based attention (RBA) local weighting to efficiently compute adaptive weights for
the training points using the residuals of their given losses, helping the optimizer to focus on high error regions. Wang et al. \cite{wang2021understanding} proposed adaptive global weights based on maximum and mean statistics of loss term gradients. The difficulty of balancing the loss terms and non-convex optimization, combined with spectral bias of the neural networks, have made application of PINNs to problems with high-frequency features and multi spatiotemporal scales limited. Addressing these issues is critical for enabling the application of PINNs to more complex and unsteady systems such as turbulence and multiphase problems. Several strategies including Fourier feature embedding \cite{song2023simulating}, and multi-scale deep neural networks \cite{XU2025113905} have been shown to improve the neural network predictions in the solution of multiple PDEs. However, application of these methods to high-dimensional PINNs and experimental noisy data is yet to be explored. Recent developments on application of second-order optimization methods have shown significantly higher robustness to spectral bias, and to the choice of loss weighting strategy in training PINNs. \cite{khodakarami2026spectral, khodakarami2025mitigating}. However, application of second-order optimizers for deep learning has been limited to low-dimensional problems due to computational cost of Hessian inverse calculations during the training of high-dimensional problems. On the other hand, quasi second-order optimizers such as SOAP \cite{vyas2024soap} approximate the second-order information in a more computationally efficient manner, making them tractable for high-dimensional problems. \cite{wang2025simulating} \\}

{In addition to the optimization scheme, the representational model also plays a role in successful training of PINNs. Choices of neural architecture and activation functions have been shown to have meaningful impact on spectral bias mitigation. \cite{khodakarami2026spectral} Recently, Kolmogorov-Arnold networks (KANs) were proposed as a new class of neural architecture \cite{liu2024kan} inspired by the Kolmogorov-Arnold representation theorem, stating that any multivariate continuous function can be expressed as a finite superposition of univariate functions and addition operations. Recent studies have demonstrated that KANs exhibit reduced spectral bias compared to MLPs, and can better learn multiscale and high-frequency functions \cite{wang2024expressiveness, zhang2025bubbleonet, yang2025multi}, making them a promising alternative to MLPs for scientific machine learning.\\}

{In this study, we take advantage of physics-informed KANs (PIKANs) \cite{toscano2025pinns, wang2025kolmogorov} to solve the 3D and unsteady heat conduction problem subjected to a highly complex boiling boundary condition that is obtained directly from high-resolution measurements of pool boiling experiments. Particularly, the spatiotemporal surface temperatures during boiling are measured through infrared (IR) thermography. Due to the chaotic and stochastic nature of nucleate boiling, vapor bubbles nucleate, grow, coalesce, and depart the surface in rapid succession, exhibiting irregular and evolving temperature fields. These bubbles appear randomly at different locations and may dynamically interact with nearby bubbles, giving rise to localized and high-frequency temperature variations in space and time. In order to accurately estimate the spatiotemporal boiling heat flux, it is necessary to accurately resolve the bubble dynamics and thermal gradients at the liquid-vapor interfaces on the boundary while solving the 3D and unsteady heat problem. Since the boundary data are obtained through experimental measurements, the model must also be robust to noise while preserving high-frequency and fine-scale features associated with physical bubble dynamics. This presents a significant challenge given the aforementioned spectral bias of neural networks, meaning they struggle to capture such sharp gradients or rapidly changing structures. As a result, high-frequency information essential for resolving bubble interfaces and locally varying heat fluxes can be severely over-smoothed. On the other hand, inherent noise in the experimental data impose a significant challenge for traditional numerical solvers that are traditionally applied to solve this problem. Even small amounts of noise in the boundary condition can propagate and become substantially amplified when computing spatial and temporal derivatives required in the heat conduction equation. Because heat conduction is governed by second-order spatial derivatives, noisy boundary values can lead to large oscillations in computed gradients, resulting in spurious and non-physical heat flux estimations.\\}

We overcome these limitations by developing a PIKAN with Chebyshev polynomials as the univariate functions for solving a time-dependent and 3D heat conduction problem with the experimentally measured boiling boundary condition. We demonstrate the importance of using quasi second-order optimizers and adaptive re-sampling to capture the high-frequency bubble dynamics. We compare PINNs and PIKANs trained with different optimizers, emphasizing the importance of both network architecture and optimization scheme. \\

The manuscript starts with a detailed definition of the problem, introduction to the experimental dataset, and description of the physics-informed network methodology in section \ref{section2}. We show the comparison of the physics-informed networks with a full numerical simulation in section \ref{section3}. In section \ref{section4}, we show the prediction results for the 3D steady and unsteady heat problem and robustness to experimental noise, followed by a summary of the methods and results in section \ref{section5}.

\section{Problem Formulation}
\label{section2}

\noindent An experimental facility is designed to investigate pool boiling heat transfer using synchronized high-speed infrared (IR) and internal-reflection-based optical imaging. A detailed description of the experimental setup and measurement technique is provided in our previous works \cite{bongarala2024method, Vydyula2026Modeling}. A brief summary of the experimental details is provided here.\\

A 1 mm thick, 25.4 mm x 25.4 mm sapphire substrate, coated with a thin indium tin oxide (ITO) layer at the center, serves as a visibly transparent and infrared-opaque heater, enabling simultaneous temperature measurement and liquid-vapor phase distribution visualization through the heater surface. Gold contact pads are deposited on the either end of the ITO region for electrical connections. The heater substrate is mounted into a pressure-controlled, saturated boiling chamber. Nucleate pool boiling experiments are conducted with HFE-7100 as the working fluid. Two immersion heaters are used to degas the fluid prior to testing, and subsequently, to maintain saturated conditions during testing. During degassing, a Graham condenser is used to prevent vapor loss while allowing the non-condensable gases to purge from the chamber. At the start of the experiment, the chamber is sealed, and a coiled condenser placed in the vapor space of the chamber regulates the chamber pressure by condensing the generated vapor. Pool temperatures are monitored using two thermocouples, and a pressure transducer records vapor pressure. Three synchronized cameras capture the boiling process. One high-speed camera (Phantom v1212), equipped with a 200 mm macro lens, captures side-view images with diffuse white LED backlighting. A second high-speed camera (Phantom VEO-710L) captures the liquid–vapor phase distribution at the heater surface using the phase-dependent reflection of red LED light source from the heater substrate. An infrared camera (ImageIR 9450hp), equipped with a 1.0x microscopic lens with a 200 mm working distance (InfraTec 3100), measures temperature distribution maps at the heater surface. Synchronization and triggering using two pulse generators ensures precise temporal frame alignment across all cameras. A DC power supply provides controlled heating to the sapphire-ITO substrate. Synchronous transient temperature and phase distribution measurements are obtained at various heat input levels along the nucleate boiling curve up to critical heat flux (CHF). The experimental dataset is available at \cite{DVN/LKAQFW_2026}
. Specifically, experimental data obtained at a steady input heat flux of 15.03 $W/cm^2$ are analyzed in this work.

\subsection{Physics-Informed Network Methodology}

\noindent PINNs and PIKANs provide a framework to solve PDEs by embedding the governing physical laws directly into the training process of the network. The solution is obtained by minimizing a composite loss function that enforces PDE residuals, initial conditions, boundary conditions, and any available experimental measurements. In this work, we develop physics-informed networks to solve a three-dimensional and unsteady heat conduction problem subject to boundary conditions obtained from the experimentally measured boiling surface temperatures. Two physics-informed architectures, including PINN and PIKAN are considered. To clarify this distinction throughout our study, PINN strictly denotes the architecture utilizing an MLP backbone, whereas PIKAN refers exclusively to the KAN backbone. While both approaches share the same physical formulation and training objective, they differ fundamentally in the functional representation of the temperature field." \\

The transient heat conduction in a solid domain $\Omega \subset \mathbb{R}^3$ with a constant thermal conductivity \textit{k} and no heat generation source is governed by the following equation, where $\rho$ is the density, $c_p$ is the specific heat capacity, $T$ is the temperature, and $T_0$ is the initial temperature condition.

\begin{equation}
\rho c_p \frac{\partial T(\mathbf{x}, t)}{\partial t}
=
k \nabla^2 T(\mathbf{x}, t),
\qquad
T(\mathbf{x},0) = T_0(\mathbf{x}),
\quad
\mathbf{x} \in \Omega,
\quad
t \in [0, t_f].
\end{equation}

\begin{figure}[H]
    \centering
    \includegraphics[width=1.1\linewidth]{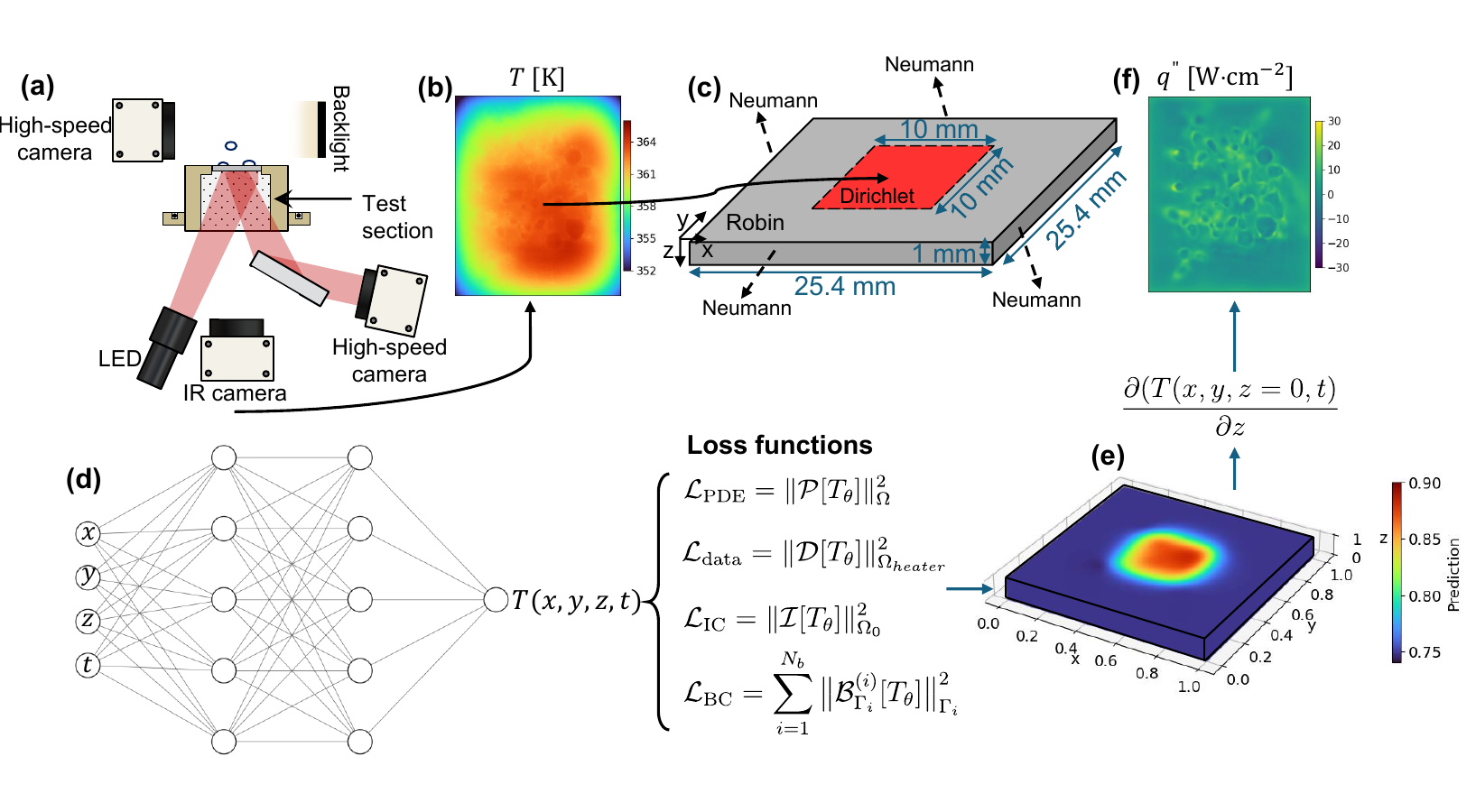}
    \caption{\textbf{Problem formulation and physics-informed netowrk (PINN and PIKAN) methodology}. (a) Schematic of experimental boiling facility. (b) Example of temperature measurements. (c) Geometry of the problem solved with PINN. The boundary conditions at each surface are noted for clarity. (d) Schematic of PINN and loss functions. (e) Example of 3D predictions by PINN. (f) Example of boiling heat flux estimation from PINN results.}
    \label{fig:Problem_formulation}
\end{figure}

\noindent All boundaries except for the upper surface exposed to the boiling fluid are assumed to be insulated. This exposed surface consists of a central heater region (ITO region) where boiling occurs and a peripheral sapphire glass region (sapphire region) where single phase cooling is happening. The experimentally measured boundary data are only available in the heater region where they are directly imposed as a data-driven Dirichlet boundary condition. A single-phase heat transfer coefficient ($h$), calibrated using natural convection measurements performed before the onset of nucleate boiling, is imposed as a Robin boundary condition (Fig. \ref{fig:Problem_formulation}d). The boundary conditions are summarized below, where $T_{exp}$ are the experimentally measured surface temperatures during boiling.

\begin{equation}
T(\mathbf{x}, t) = T_{\mathrm{exp}}(\mathbf{x}, t),
\qquad
\mathbf{x} \in \partial \Omega_{\mathrm{ITO}},
\quad t \in [0, t_f]
\end{equation}

\begin{equation}
-k \nabla T(\mathbf{x}, t) \cdot \mathbf{n} 
= h \left( T(\mathbf{x}, t) - T_\infty \right),
\qquad
\mathbf{x} \in \partial \Omega_{\mathrm{sapphire}},
\quad t \in [0, t_f]
\end{equation}

\begin{equation}
-k \nabla T(\mathbf{x}, t) \cdot \mathbf{n} = 0,
\qquad
\mathbf{x} \in
\partial \Omega \setminus
\left(
\partial \Omega_{\mathrm{ITO}} \cup \partial \Omega_{\mathrm{sapphire}}
\right),
\quad t \in [0, t_f]
\end{equation}

The PINN uses an MLP architecture, where each node at each layer performs a linear transformation of all the outputs of the previous layer via weights ($W$) and biases ($b$), followed by a non-linear transformation using an activation function. The conventional MLP operations are shown in the following equation.

\begin{equation}
T_\theta(\mathbf{x}, t)
=
\mathbf{W}^{(n)}
\;\sigma^{(n-1)}\Big(
\mathbf{W}^{(n-1)}
\;\sigma^{(n-2)}\Big(
\cdots
\;\sigma^{(0)}\big(
\mathbf{W}^{(0)}
(\mathbf{x}, t) 
+ \mathbf{b}^{(0)}
\big)
+ \mathbf{b}^{(1)}
\cdots
\Big)
+ \mathbf{b}^{(n-1)}
\Big)
+ \mathbf{b}^{(n)},
\label{eq:MLP}
\end{equation}

\noindent where, $\sigma^{(i)}$ represents the activation function at the \textit{i}-th layer. Although activation functions are usually constant within the network, here we use layer-wise adaptive activation functions that are trained concurrently with the weights and biases. MLPs are universal function approximators, meaning they can learn any continuous mapping given enough width and depth. In this work, we use a network depth of 8 layers and width of 240, with layer-wise adaptive Swish \cite{JAGTAP2020109136} as the activation function. To help the PINN with capturing the high-frequency features, a Fourier feature embedding layer\cite{tancik2020fourier} was used before the first layer of the network. \\

The PIKAN uses a KAN as the backbone architecture. KANs are insprired by Kolmogovo-Arnold representation theorem stating that any multivariate continuous function can be written as composition of univariate functions:

\begin{equation}
f(x) = \sum_{m=1}^{M} \Phi_m \Bigg( \sum_{n=1}^{N} \psi_{mn}(x_n) \Bigg),
\label{eq:KAN_simple}
\end{equation}

\noindent where $\psi_{mn}$ and $\Phi_m$ are univariate functions \cite{liu2024kan}. In MLPs, fixed (or adaptive) activation functions are used on nodes and learnable parameters are trained across the edges. However, in KANs, learnable activation functions (e.g., univariate functions) are learned on edges and only summation operation is done on nodes. Therefore, KANs are formulated as follows:

\begin{equation}
f(\mathbf{x}) \approx
\sum_{i_{L-1}=1}^{n_{L-1}} 
\phi_{L-1, i_L, i_{L-1}} \Bigg(
\sum_{i_{L-2}=1}^{n_{L-2}} 
\cdots
\Big(
\sum_{i_2=1}^{n_2} 
\phi_{2, i_3, i_2} \Big(
\sum_{i_1=1}^{n_1} 
\phi_{1, i_2, i_1} \Big(
\sum_{i_0=1}^{n_0} 
\phi_{0, i_1, i_0}(x_{i_0})
\Big)
\Big)
\Big)
\Bigg),
\end{equation}

\noindent where $L$ is the number of layers, $n_j$ is the number of neurons in the \textit{j}-th layer, and $\phi_{k,j,i}$ are the univariate activation functions \cite{shukla2024comprehensive}. Here, we used Chebyshev polynomials as univariate functions to construct our KAN architecture. Although several types of univariate functions such as spline, radial basis function (RBF), Wavelets, Jacobi polynomials, and Chebyshev polynomials have been proposed to be used in KANs, Chebyshev polynomials have shown more success in solving problems involving high frequencies \cite{shukla2024comprehensive, toscano2025aivt}. To ensure that the input to each Chebyshev polynomial remains within the standard [-1, 1] interval, a tanh activation is applied before each layer.\\

For either of the networks denoted as $N_{\theta} (x,y,z,t)$, the training objective shown in the following equation minimizes a weighed sum of five loss terms including PDE residuals, initial conditions, insulated boundary condition at all sides except for the boiling surface, a Dirichlet boundary condition at the heater region, and a Robin boundary condition at the sapphire region surrounding the heater (Fig. \ref{fig:Problem_formulation}c).

\begin{equation}
\mathcal{L} =
\lambda_{\mathrm{PDE}} \, \mathcal{L}_{\mathrm{PDE}}
+ \lambda_{\mathrm{IC}} \, \mathcal{L}_{\mathrm{IC}}
+ \lambda_{\mathrm{BC}}^{\mathrm{Dirichlet}} \, \mathcal{L}_{\mathrm{BC}}^{\mathrm{Dirichlet}}
+ \lambda_{\mathrm{BC}}^{\mathrm{Robin}} \, \mathcal{L}_{\mathrm{BC}}^{\mathrm{Robin}}
+ \lambda_{\mathrm{BC}}^{\mathrm{Neumann}} \, \mathcal{L}_{\mathrm{BC}}^{\mathrm{Neumann}},
\label{eq:Loss}
\end{equation}

\noindent where $\lambda_{\mathrm{PDE}}, \lambda_{\mathrm{IC}}, \lambda_{\mathrm{BC}}^{\mathrm{Dirichlet}}, \lambda_{\mathrm{BC}}^{\mathrm{Robin}}$, and $\lambda_{\mathrm{BC}}^{\mathrm{Neumann}}$ are pre-defined fixed weights that are modified using a pre-defined scheduler. The scheduler assigns larger weights to the $\lambda_{\mathrm{BC}}^{\mathrm{Dirichlet}}$ at the start of the training due to the complexity of the experimental data. Then, it sequentially increases the weights for $\lambda_{\mathrm{IC}}$, $\lambda_{\mathrm{BC}}^{\mathrm{Robin}}$, $\lambda_{\mathrm{BC}}^{\mathrm{Neumann}}$, and $\lambda_{\mathrm{PDE}}$ enabling a sequential learning strategy. Sequential learning eases the optimization by decomposing the multi-objective and highly coupled optimization into simpler subproblems. Instead of enforcing all loss terms from the beginning of the training, the network first learns an initial parameter space by fitting experimental data and satisfying other boundary conditions. This provides a good initialization in parameter space, reducing the chance of getting trapped in a poor local minima early in the training stage. Sequentially introducing the loss terms improves conditioning of the optimization problem, and reduces the competition between loss components. As a result, this can make the convergence faster, especially when combined with adaptive weighting strategies.

\subsection{Training and Optimization Strategies}
Recent application of second-order optimization for PINNs have shown promising results in solving complex PDEs and mitigating spectral bias. Most quasi-Newton methods such as BFGS or its low-memory variant, L-BFGS, and self-scaled Broyden (SSBroyden) \cite{KIYANI2025118308, urban2025unveiling} have shown multiple orders of magnitude enhancements compared to the conventional Adam optimizer. However, their application has been mainly limited to full-batch training and they suffer from scalability to high-dimensional problems. In batch training, the mini-batch changes at each iteration, potentially causing instability in these methods. This is because these methods depend on gradient differences to update Hessian approximations, and the variability in data across batches can introduce inconsistency in those gradients. Therefore, their application has remained limited to low-dimensional problems where full-batch training is feasible. Another recently developed optimizer is SOAP \cite{vyas2024soap} which is a quasi-second-order optimization method that has shown superior performance over Adam in mini-batch training for high-dimensional problems such as training of language models \cite{vyas2024soap} and PINNs for turbulence simulation \cite{wang2025simulating}. In SOAP, the optimizer modifies the gradient matrix ($G$) by leveraging the structural information of the Jacobian matrix through a spectral lens. Two pre-conditioner matrices, $L^{t} = \beta L^{t-1} + (1-\beta) G G^T$ and $R^{t} = \beta R^{t-1} +  (1-\beta) G^T G$ are constructed from $G$, where $\beta$ is a forgetting factor balancing the memory of the past gradient geometry and the new update. By performing eigen-decomposition on these pre-conditioners,  the principal directions of the gradient geometry are identified. Then, $G$ is projected into this eigenspace, and Adam steps are conducted in this space ($G_{rotated}$).

\begin{equation}
    L = Q_L \Sigma_L Q_L^\top, \quad R = Q_R \Sigma_R Q_R^\top
    \label{eq:eigendecomposition}
\end{equation}

\begin{equation}
    G_{\text{rotated}} = Q_L^\top G Q_R
\end{equation}

\noindent By applying Adam updates on $G_{rotated}$ instead of the original parameter space, the optimizer effectively performs adaptive steps in a coordinate system where the parameters are decorrelated. The update is then projected back to original space ($\Delta W_t$), allowing for more efficient navigation of the loss landscape and potentially more robustness to avoid early stagnation in local minima compared to standard first-order methods such as Adam.

\begin{equation}
    \hat{G}_t = \text{Adam}(G_{\text{rotated}, t}, \beta_1, \beta_2, \epsilon)
\end{equation}

\begin{equation}
    \Delta W_t = Q_L \hat{G}_t Q_R^\top
\end{equation}

\noindent New parameters in the original space (($W_{t+1}$)) are calculated using the update ($\Delta W_t$), and learning rate ($\eta$) as shown below.

\begin{equation}
    W_{t+1} = W_t - \eta \Delta W_t
\end{equation}

Here, we employed the SOAP optimizer in a mini-batch training scheme to develop both the PINN and PIKAN models. In addition to using a quasi-second-order optimization method, we also developed a residual-based adaptive resampling method during the training. To avoid instability during early training, training starts with random sampling for the first 10$\%$ - 20$\%$ of the total training iterations. Subsequently, residual-based resampling for the Dirichlet boundary as well as for the collocation points used for the PDE loss calculation is imposed; however, random sampling was retained for the other boundaries and initial condition throughout the training. The objective of residual-based resampling is to preferentially resample with higher probabilities from high-residual regions, thereby directing the optimization to focus on those hard-to-learn regions. These regions often correspond to locations with high frequency features or sharper gradients (e.g., at bubble interfaces). A temperature parameter (note this parameter is different than the physical temperature discussed elsewhere in this work.) was introduced to control the sharpness of the sampling distribution, where higher temperature resulted in more uniform sampling probability, whereas lower temperatures concentrated sampling more strongly around high-residual regions.\\

Let $\{ r_i \}_{i=1}^{N}$ denote the residual values evaluated at $N$ candidate sampling points. 
A temperature-scaled residual-based probability distribution is first defined as
\begin{equation}
p_i^{\mathrm{res}} =
\frac{\exp\!\left( r_i / \tau \right)}
{\sum_{j=1}^{N} \exp\!\left( r_j / \tau \right)},
\qquad i = 1, \dots, N,
\label{eq:residual_softmax}
\end{equation}
where $\tau > 0$ is a temperature parameter controlling the sharpness of the distribution. To prevent excessive concentration of sampling points and to maintain sufficient exploration, 
the residual-based distribution is blended with a uniform distribution
\begin{equation}
p_i^{\mathrm{uni}} = \frac{1}{N}.
\label{eq:uniform_sampling}
\end{equation}

\noindent The final sampling probability is obtained via combination of the two distributions:
\begin{equation}
\tilde{p}_i =
\alpha \, p_i^{\mathrm{res}}
+
(1 - \alpha)\, p_i^{\mathrm{uni}},
\qquad 0 \le \alpha \le 1,
\label{eq:blended_sampling}
\end{equation}
where $\alpha$ controls the relative importance of residual-based sampling ($\alpha$ = 0.8 in the study). For the residual-based resampling, setting $\tau$ to a very small value (e.g., 0.1) will make sure that the sampling is highly skewed toward large residual regions. However, this might also cause instability and failure in the training if done very early. Our experiments demonstrated that gradually decreasing $\tau$ throughout the training provides the best trade-off between stability and capturing high-frequency features resulting in further minimization of the loss function. Therefore, we used the following strategy to gradually change $\tau$ during the training:
\begin{equation}
    \tau = max(\tau_{low}, 0.9^{(n - n_{resample})}),
    \label{eq:Tau}
\end{equation}

\noindent where $n$ is the training iteration, $n_{resample}$ is the iteration from which the probabilistic resampling starts, and $\tau_{low}$ is the minimum value of $\tau$ which is set to 0.2 in this work.

\section{Comparison with Numerical Simulation}
\label{section3}

\begin{figure}[H]
    \centering
    \includegraphics[width=1.1\linewidth]{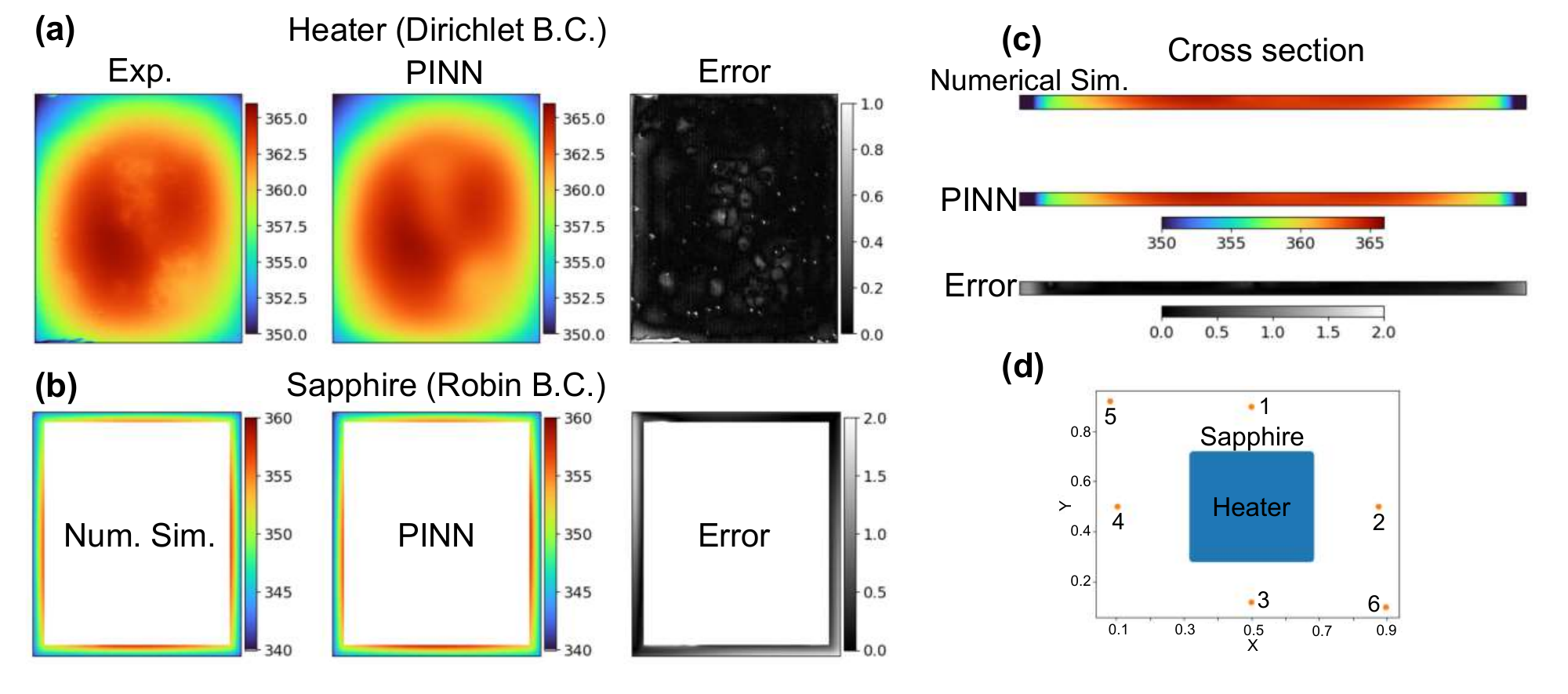}
    \caption{\textbf{Comparisonof PINN results without synthetic sensor measurements on the Sapphire region with numerical simulation data.} (a) Predictions on the heater region (Dirichlet B.C.). (b) Predictions on the sapphire glass region (Robin B.C.). (c) Predictions on a cross-section in the middle of the geometry. (d) Schematic showing location of heater and sapphire regions, and potential synthetic sensors.}
    \label{fig:PINN_validation}
\end{figure}

\noindent The PINN solution is first validated against a numerical simulation with the results as shown in Fig. \ref{fig:PINN_validation}. 
{
The numerical simulations are based on the finite volume method (FVM) and solved using Ansys Fluent by implementing a user-defined function to impose the experimentally measured temperature boundary condition. A steady-state 3D heat conduction simulation was first performed with the first time-step surface temperature distribution to obtain an initial temperature distribution estimate in the substrate. This temperature field was then used to initialize the unsteady 3D simulation with the experimentally measured spatiotemporally varying temperature boundary condition. Additional details of the numerical simulation procedure are provided in our previous work \cite{bongarala2024method}.
}

The PINN accurately predicts temperatures across different regions of the domain. The largest errors occur at the sapphire region surrounding the heater region, where the Robin boundary condition is imposed. This increased error is expected because accurate satisfaction of the Robin boundary condition requires both the temperature field and its spatial derivatives to be estimated correctly. Table \ref{tab:mae_relerr} summarizes the mean absolute error (MAE) and relative error for cases with and without synthetic sensor measurements on the Robin boundary. Note that using such synthetic sensor measurements (i.e., providing the numerical simulation values at these location as additional data losses during PINN training) enhances the predictions on the Robin boundary region, but has limited impact on the other regions. These results are based on the Adam optimizer. Although the overall temperature predictions are accurate, the PINN fails to capture the localized bubble signatures within the heater temperature region (Fig. \ref{fig:PINN_validation}a).

It should be noted that the difference as compared to the heat fluxes obtained by numerical simulations are shown as an error as a standard practice, these should not necessarily be considered as the ground truth or exact solution due to the highly noisy estimations that result from nonphysical noise in the temperature data. Nevertheless, they can be used as a soft and qualitative validation of the PINN results. \\

\begin{table}[h!]
\centering
\caption{PINN mean absolute error (MAE) and relative error for different regions under varying synthetic sensor configurations.}
\begin{tabular}{l|cc|cc|cc}
\hline
\multirow{2}{*}{Region} & 
\multicolumn{2}{c|}{0 Sensors} & 
\multicolumn{2}{c|}{4 Sensors} & 
\multicolumn{2}{c}{6 Sensors} \\ \cline{2-7}
 & MAE & Rel. Err. & MAE & Rel. Err. & MAE & Rel. Err. \\ \hline
Heater & 0.0783 & 0.0217\% & 0.0755 & 0.0210\% & 0.0692 & 0.0192\% \\
Sapphire & 0.767 & 0.220\% & 0.313 & 0.0893\% & 0.291 & 0.0830\% \\
Domain Interior & 0.197 & 0.056\% & 0.111 & 0.0313\% & 0.104 & 0.0294\% \\ \hline
\end{tabular}
\label{tab:mae_relerr}
\end{table}

\section{Results and Discussion}
\label{section4}

\subsection{Steady-State Results}

\begin{figure}[H]
    \centering
    \includegraphics[width=1.0\linewidth]{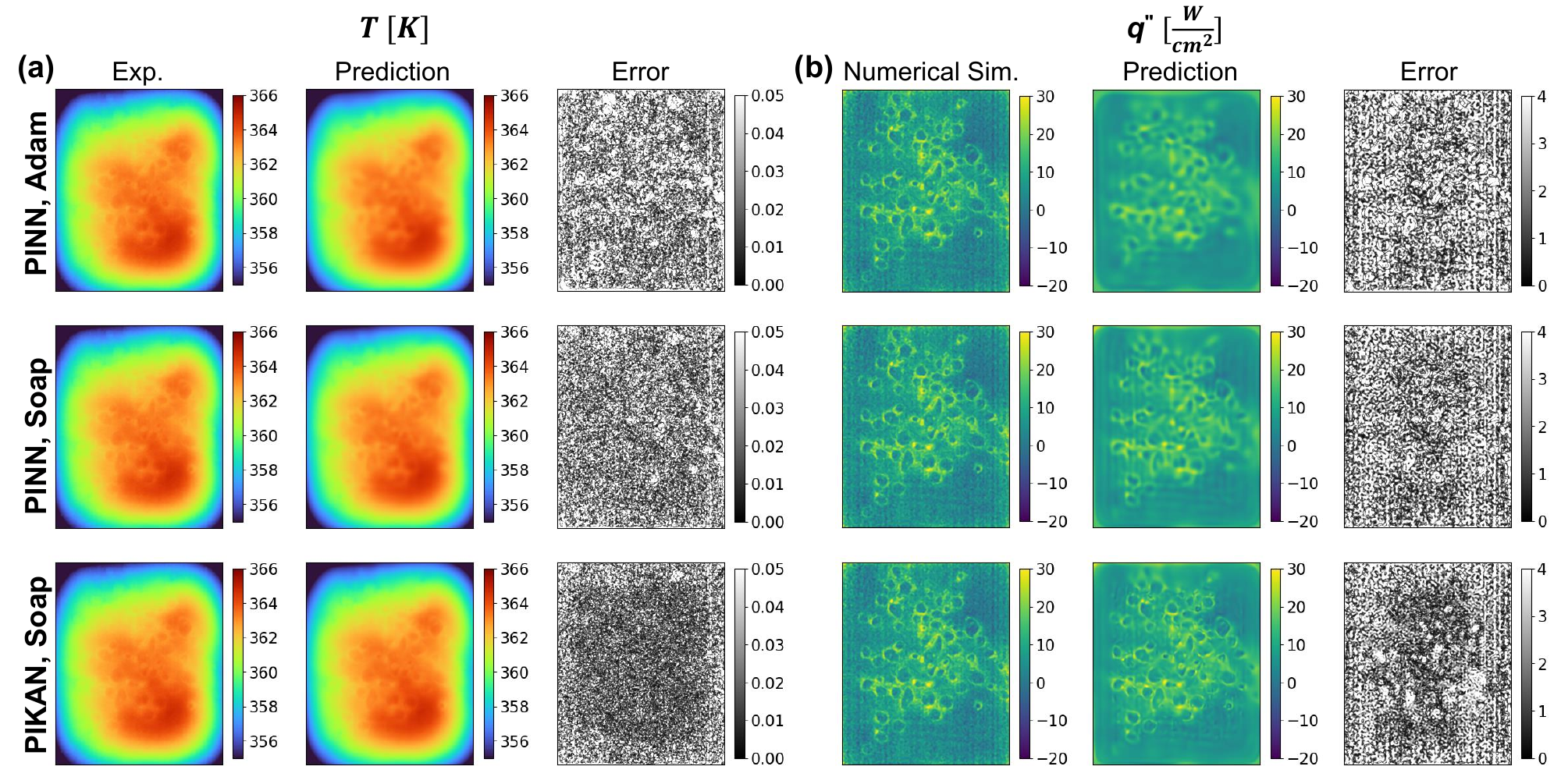}
    \caption{\textbf{Physics-informed networks predictions on the heater region for the steady-state problem.} (a) Temperature predictions by PINN with Adam and SOAP optimizers, and PIKAN with SOAP optimizer. Exp.: experimental temperature measurements on the heater region. (b) Heat flux ($q^"$) estimations by PINN with Adam and SOAP optimizers, and PIKAN with SOAP optimizer. Numerical simulation (Numerical Sim.) results are shown for comparison. These results show the boiling heat flux (heat flux from the surface to the fluid).}
    \label{fig:steady_state}
\end{figure}

\noindent We first solve the steady-state heat equation and use the resulting temperature field as the initial condition to solve the transient simulation. Fig. \ref{fig:steady_state} shows the steady-state temperature and heat flux predictions over the heater region with PINN and PIKAN, trained with Adam or SOAP optimizers. The predicted 3D temperature fields were used to estimate the heat flux at the heater surface that can be used to estimate the spatiotemporal boiling heat flux. While the heat fluxes predicted by the simulation indicate the amount of heat going from the heater into the substrate, the boiling heat fluxes (Fig. \ref{fig:steady_state}b) are estimated by subtracting this substrate heat flux from the total heat flux imposed on the heater.\\

 The predicted temperatures for PINN optimized with Adam looks slightly oversmoothed compared to the experimental data. However, the oversmoothing is worsened in the  heat flux prediction, where many of the interfaces observable in the numerically simulated results are not properly captured by the PINN (Adam), and the errors are  higher in regions where bubbles exist. By simply changing the PINN optimizer to SOAP, the predictions are slightly improved; however, the predictions still suffer from oversmoothing, especially at the liquid vapor interfaces of bubbles where there are sharp gradients in the temperature field due to high local heat flux dissipation at the contact line. This is due to the spectral bias of the PINN architecture (namely, MLP). Adding a Fourier feature embedding to the first layer could not resolve this problem. However, when using PIKAN with SOAP optimizer, it is apparent that the predictions are enhanced at bubble regions, as most of the interfaces appear to be more sharply captured. These results demonstrate the advantage of a KAN architecture in spectral bias mitigation for a high-dimensional problem. 

\begin{figure}[H]
    \centering
    \includegraphics[width=1.0\linewidth]{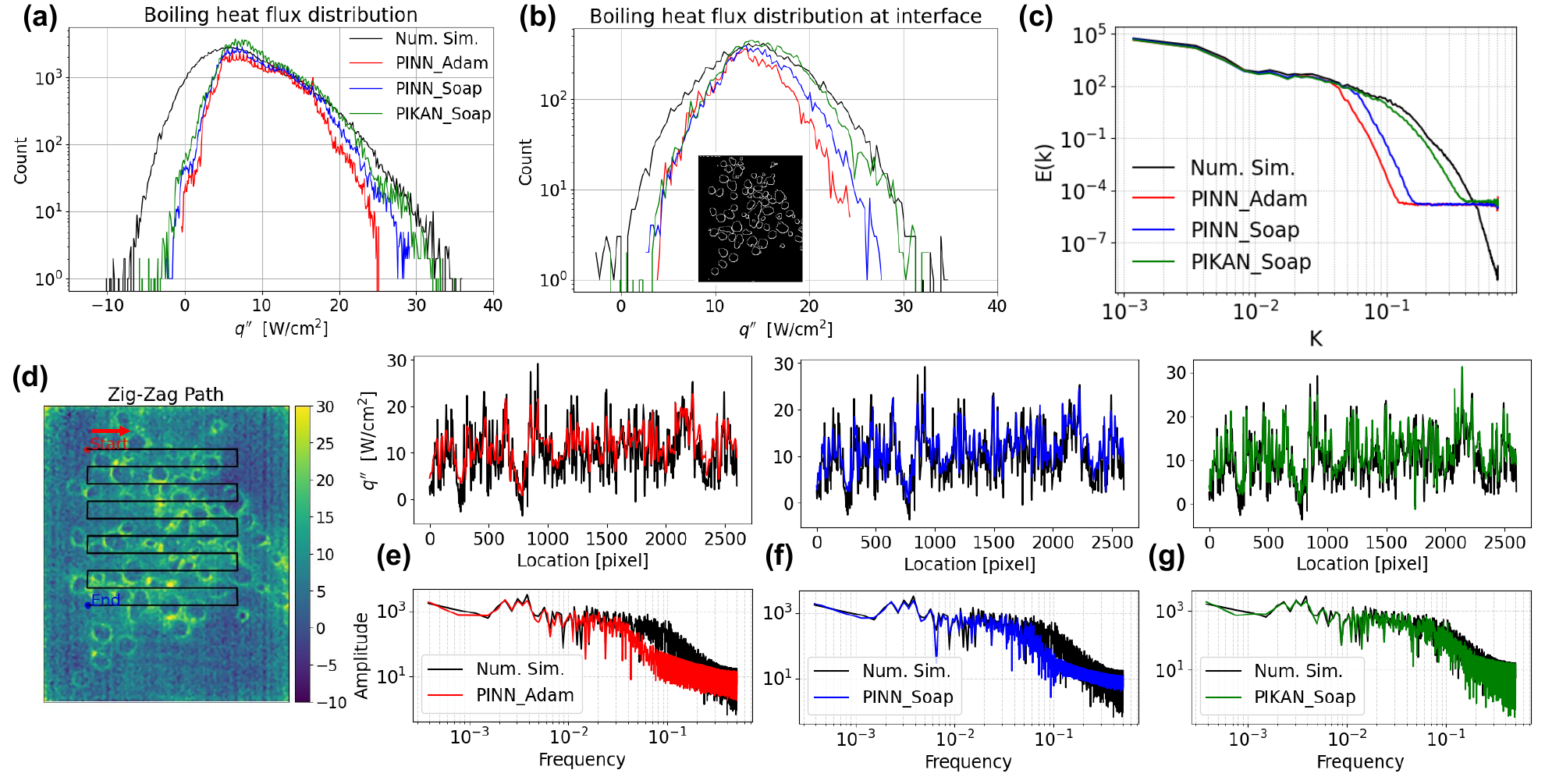}
    \caption{\textbf{Analysis of physics-informed networks predictions for the steady-state problem.} (a) Heat flux distribution on the heater. (b) Heat flux distribution at the bubbles interfaces. (c) Spatial energy spectrum plot for each of the estimations. (d) Manual zig-zag path chosen on the heater region (from "Start" to "End"). (e-g) Heat flux estimation and Fourier amplitude spectrum on the zig-zag path for (e) PINN with Adam, (f) PINN with SOAP, and (g) PIKAN with SOAP. Num. Sim.: numerical simulation}
    \label{fig:Steady_State_Analysis}
\end{figure}

The experimentally measured temperatures contain a small amount of noise that is propagated into the numerical solver, resulting in large unphysical fluctuations in the heat flux estimations with the numerical method. However, the PIKAN correctly avoids the inherent experimental noise while capturing the bubble interfaces sharply. This is an important advantage of the PIKAN over the numerical simulation. Fig. \ref{fig:Steady_State_Analysis}a-b demonstrates the heat flux distribution on heater region and only at the bubble interfaces, respectively. The bubble interfaces were defined using the corresponding synchronized phase distribution maps.  The binarized phase maps were dilated to generate a two-pixel-wide outward band surrounding the phase boundary, which was defined as the effective interface. The PIKAN optimized with SOAP can capture the right tail of the distribution which corresponds to the positive high heat fluxes dissipated into the fluid at the bubble interfaces. Similarly, the energy spectrum results (Fig. \ref{fig:Steady_State_Analysis}c) demonstrates that significantly large high-frequency energies are recovered with PIKAN trained with SOAP. However, PINN optimized with SOAP and either of PINN or PIKAN trained with Adam fail to recover most of the energy at high wavenumbers, resulting in the loss of information at bubble interfaces. To further investigate this difference between the model predictions, the results in Fig. \ref{fig:Steady_State_Analysis}(d-g) show the heat flux estimations in a zigzag line passing through the bubbly region. The estimations by the PINN trained with Adam (Fig. \ref{fig:Steady_State_Analysis}e) entirely misses the peaks and large spatial variations of the boiling heat flux. The results are slightly improved by switching the optimizer from Adam to SOAP (Fig. \ref{fig:Steady_State_Analysis}f), however, it still exhibits significant spectral bias to low-frequencies and the energy of high wavenumbers is underestimated. On the other hand, the result of the PIKAN trained with SOAP captures the high-frequency spatial variation of the heat flux much more accurately as further confirmed in the Fourier spectrum of the estimation (Fig. \ref{fig:Steady_State_Analysis}g). Note that, other than the neural architecture and the optimizer, all the other training hyperparameters as well as the number of parameters in each model were kept the same.\\

\subsection{Unsteady Results}

\noindent The PIKAN predictions for the steady-state case were used as the initial condition for solving the unsteady problem with a time-step of 2 ms. Similar to the steady case, a sequential learning strategy was used to facilitate the optimization. The learning started at the boiling boundary and the initial condition, followed by other boundaries and the PDE loss term. Near the end of optimization, smaller weights were used for the initial condition and weight for the PDE loss was increased. Given the dimensionality of the problem, full-batch training was not a tractable option due to GPU memory constraints. Instead we used mini-batching during the training. In the first 20\% of the iterations, random sampling was used to take 10\% of the collocation points at every iteration. After this time, we used adaptive resampling as shown in Eqs. \ref{eq:residual_softmax} - \ref{eq:Tau} to selectively choose collocation points from highly erroneous regions. Specifically, we resampled 5\% of the points in the Sapphire region around the heater region with Robin boundary condition, 5\% of the internal points inside the domain, 20\% of the points on the heater region, and 50\% of the points on the other boundary regions and initial condition points. Note that the ratios are selected based on ablation studies and ensuring that the total number of points fit on a single GPU during the training. However, the successful training was not overly sensitive to these ratios and various combinations resulted in similar performance.\\

\begin{figure}[H]
    \centering
    \includegraphics[width=1.0\linewidth]{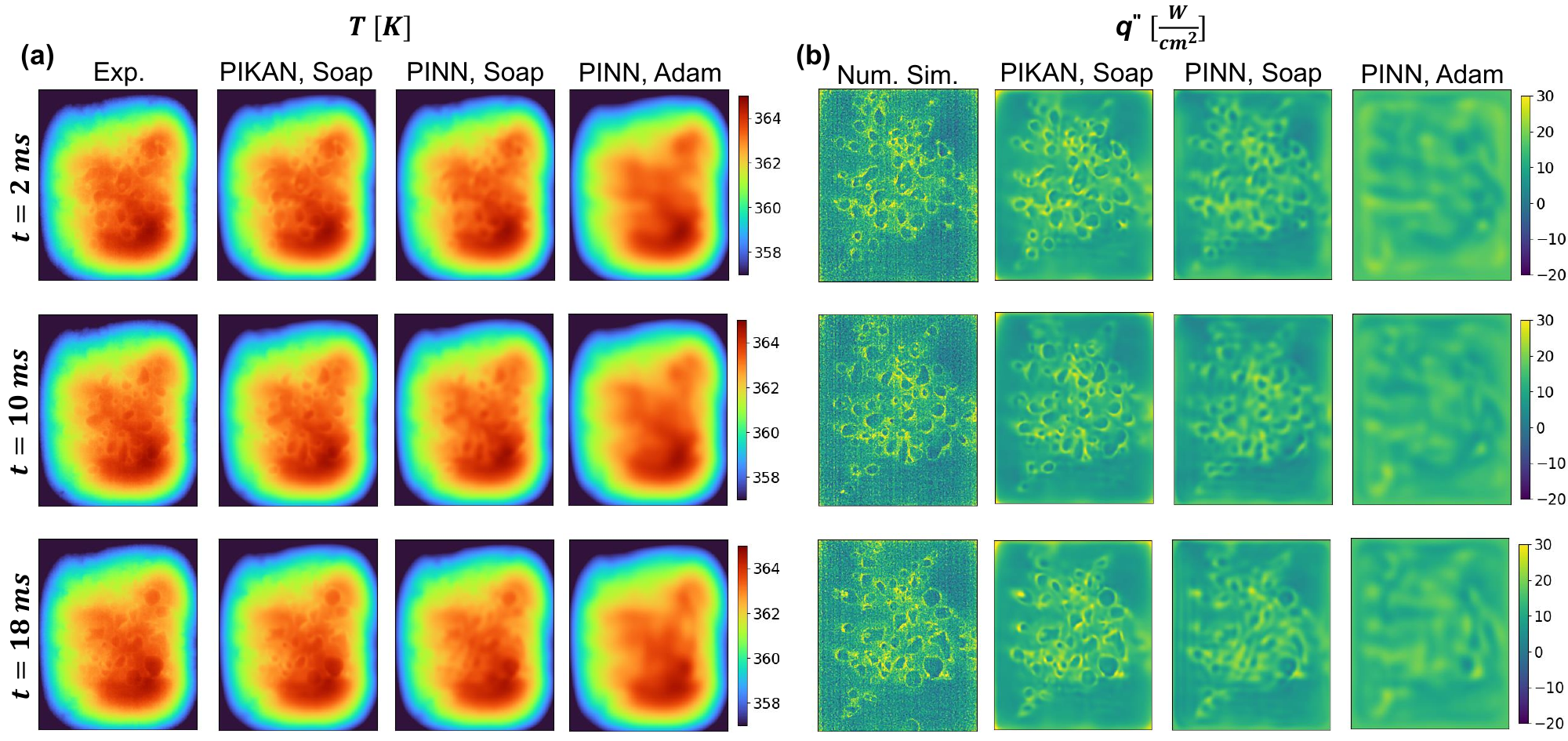}
    \caption{\textbf{Physics-informed networks predictions on the heater region for unsteady problem.} (a) Temperature predictions. (b) Boiling heat flux estimations. In each sub-figure, the first, second, and third rows correspond to different time-steps. Columns represent results of experimental temperature measurements (Exp.) in (a) or heat flux ($q^"$) estimation in (b), based on numerical simulation (Num. Sim.), PIKAN with SOAP, PINN with SOAP, and PINN with Adam from left to right. $q^"$ shows the boiling heat flux (heat flux to the fluid).}
    \label{fig:Unsteady}
\end{figure}

 Figure \ref{fig:Unsteady} shows the unsteady predictions of the heater temperatures and heat fluxes to the fluid. Note that the transient PINN field predictions using the Adam optimizer are significantly over-smoothed compared to the steady results. This is due to the additional dimension (time) in the unsteady problem which makes the optimization more challenging. Similar to the trend in steady predictions, switching the PINN optimizer to SOAP can significantly enhance the ability of the predictions to capture sharper gradients at bubble interfaces. The PIKAN, owing to a higher representational power for high-frequencies, trained with SOAP, achieves the best performance and successfully resolves the transient evolution of bubble interfaces. Videos of the predictions can be found in supplemental files. The importance of resolving the interfaces is more significant in the unsteady case. The PINN optimized with Adam shows little variation across different time-steps, illustrating that it has converged to a single temporally averaged prediction rather than capturing the bubble dynamics over time. Therefore, it also fails to capture the heat flux variations over time. On the other hand, PIKAN with SOAP accurately predicts bubble interfaces at each time-step, thus capturing the dynamics correctly. The heat flux distributions (Fig. \ref{fig:Transient_Analysis}a) shows that, except for PIKAN trained with SOAP, the rest of the models significantly fail to capture the right tail of the distribution corresponding to the high heat flux dissipated at the bubble interfaces. On the other hand, the left tail corresponds to the noisy (unphysical) heat flux estimations by the numerical simulation that are correctly filtered out by the PIKAN model. Fig \ref{fig:Transient_Analysis}b shows the heat flux distribution at a single time-step at the bubble interface locations, suggesting that the PIKAN optimized by SOAP is most effective at capturing the distribution tails.

\subsection{PIKAN Robustness to Experimental Noise}
\begin{figure}[H]
    \centering
    \includegraphics[width=1.0\linewidth]{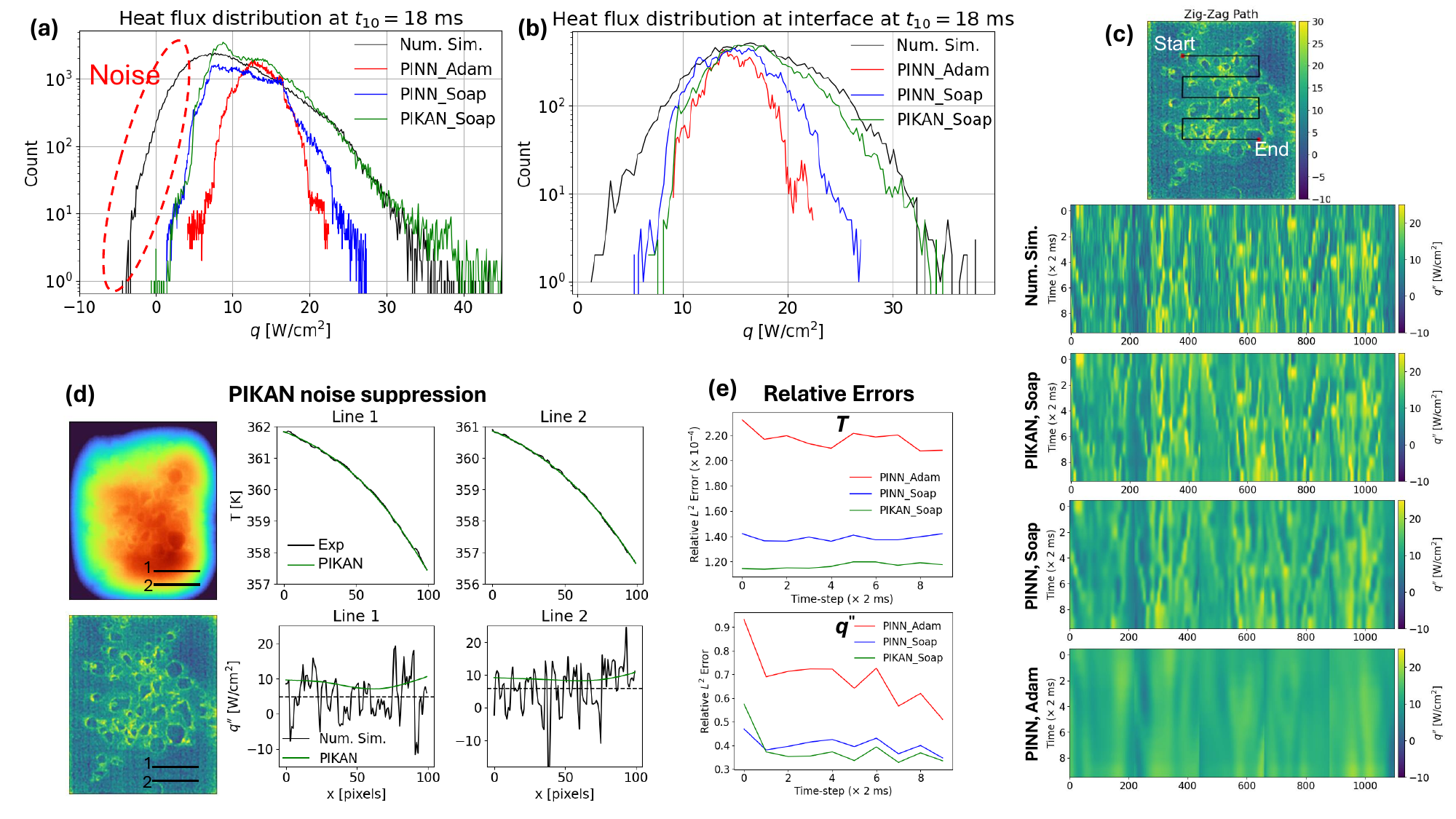}
    \caption{\textbf{Analysis of physics-informed Kolmogorov Arnold network predictions for unsteady problem.} (a) Boiling heat flux distribution at final time-step. (b) Boiling heat flux distribution at the interface at final time-step. (c) Heat flux estimations on a manual path in the bubble regions by numerical simulation (Num. Sim.), PIKAN with SOAP, PINN with SOAP, and PINN with Adam. (d) PIKAN noise suppression. (e) Relative errors of temperature (\textit{T}) prediction and heat flux (\textit{$q^"$}) estimation.}
    \label{fig:Transient_Analysis}
\end{figure}

\noindent High-speed, high-resolution IR measurements typically contain a certain level of noise within the signal \cite{bucci2016mechanistic}, which can significantly affect the estimation of derived quantities such as heat flux. Figure \ref{fig:Transient_Analysis}d shows the temperature and heat flux on two separate lines outside the region having bubble interfaces, where no physical fluctuations are expected. There is only a small amount of noise observed in the experimentally measured temperatures along these lines. However, these small perturbations are propagated by the numerical solver, affecting the 3D temperature reconstruction. Because heat flux depends on spatial temperature gradients, numerical differentiation from the solved 3D temperature distribution further magnifies measurement noise, leading to large unphysical oscillations in the reconstructed heat flux field.\\

Note that the finite element method (FEM) based on a weak formulation may be less sensitive to the measurement noise, however, they face an inherent trade-off between suppressing the noise and preserving the localized high-frequency thermal gradients associated with bubble interfaces. In contrast, PIKAN is robust to noise while preserving the physically meaningful high-frequency features in both temperature predictions and heat flux estimations. PIKAN noise suppression is shown in Figure \ref{fig:Transient_Analysis}d. The temperature predicted by the PIKAN along the two lines eliminates the small perturbations seen in experimental measurements (Fig. \ref{fig:Transient_Analysis}d), leading to stable and physically consistent heat flux estimation. Importantly, this denoising behavior does not suppress thermal signatures associated with bubble dynamics. Instead, PIKAN selectively retains the sharp temperature gradients generated near the interfaces, as previously highlighted in \ref{fig:Transient_Analysis}g, which are responsible for localized heat transfer enhancement. The impact of noise suppression is further reflected in the heat flux probability distribution (Fig. \ref{fig:Transient_Analysis}a-b). The negative (left) tail of this distribution is primarily caused by noise-induced oscillations in the numerical simulation-based heat flux estimation.\\

By removing these nonphysical fluctuations, the PIKAN effectively eliminates the left tails while fully preserving the positive (right) tail associated with large heat flux regions near bubble interfaces. This observation indicates that the model is capable of distinguishing between measurement noise frequencies and physically meaningful high-frequency phenomena, thereby improving the reliability of heat flux estimation, making it superior to the numerical simulation results. Fig \ref{fig:Transient_Analysis}e shows the relative $L^2$ for temperature predictions and heat flux estimations. Both PINN and PIKAN optimized with SOAP show significantly lower errors compared to PINN optimized with Adam. Additionally, the PIKAN yields lower errors compared to PINN, when the same SOAP optimizer is used for training them. It should be noted that due to the inherent noisy infrared measurements and large impact of noise to the numerical simulation heat flux estimation, these measures may not be the true representative of the errors, however, they still provide comparative insights for different representational models and optimizers.\\

Fig. \ref{fig:Transient_Analysis}c further examines the spatial variation of heat flux along a zigzag line traversing in the bubbly region for one of the time-steps ($t$ = 18 ms). It can be seen that the PINN optimized with Adam is completely oversmoothing the heat fluxes. However, the PIKAN optimized with SOAP can recover most of the heat flux high-frequency variations. The PINN optimized with SOAP lies in between the PIKAN-SOAP and PINN-Adam results, where it captures some of the heat flux variations in time and space, but remains over-smoothed compared to PIKAN-SOAP results.  The results demonstrate that both the neural architecture (KAN) and the optimizer (SOAP) play an important role in capturing the spatiotemporal heat flux variation .

\section{Summary}
{
\noindent In this study, we developed a PIKAN trained with a quasi-second order optimizer to solve a unsteady and 3D heat conduction problem subject to an experimentally measured boiling boundary condition. The imposed boundary condition contains spatiotemporal high-frequency features generated due to rapid bubble nucleation, growth, and departure, as well as inherent measurement noise, making the problem particularly challenging for conventional PINNs.\\

Due to the inherent spectral bias of the neural networks, PINNs trained with Adam optimizer results in significantly over-smoothed predictions, and fails to capture the localized high heat flux regions near the bubble interfaces. Our results show that both the neural representation and the optimization method play significant roles in mitigating spectral bias and improving the reconstruction of the interface-driven thermal features. By replacing the Fourier feature enhanced MLP used in the PINN with a Chebyshev PIKAN, and employing the SOAP quasi-second order optimizer, the model is able to accurately resolve the temperature gradients at the interfaces, preserve high-frequency spatial and temporal features, and capture the complex heat flux dynamics associated with boiling under unsteady operating conditions.\\

The experimental IR measurements from boiling experiments contains a certain and finite level of noise. Although the temperature fluctuations at the boundary associated with this noise are relatively small, they can be significantly amplified in the estimated heat flux fields obtained from discretized numerical solvers (e.g., FVM or FEM), resuling in unphysical oscillations in predicted surface heat fluxes induced by noisy boundary measurements. In contrast, the proposed PIKAN model demonstrates strong robustness to experimental IR measurement noise. The model selectively suppresses small nonphysical perturbations in the experimental temperature data, while preserving the physical high-frequency features associated with bubble dynamics. Therefore, it accurately resolves localized high heat flux regions near the interfaces while simultaneously eliminating the spurious heat flux oscillations observed in the numerical solver results.\\

These results suggest that our framework can potentially serve as a new infrared data-processing tool, substituting the conventional numerical methods, in solving the backside heat problem with boiling boundary conditions, as well as other measurement data, resulting in more physical heat flux estimations. Future works can enhance the computational speed of the PIKAN training for this problem using more efficient optimiziation methods and parallel computing, providing additional advantage over conventional numerical solvers.  Overall, the proposed method combines the expressive representation capability of KANs with improved optimization characteristics of quasi-second-order optimization strategy, enabling accurate reconstruction of multiscale thermal phenomena from experimental measurements subject to noise. These results demonstrate a successful application of PIKANs as a powerful tool for solving high-dimensional problems, involving complex interface dynamics with real experimental measurements, and high-frequency spatiotemporal characteristics.
}

\label{section5}

\section*{Declaration of competing interest}
\noindent The authors declare that they have no known competing financial interests or personal relationships that could have appeared to influence the work reported in this paper.

\section*{Acknowledgements}
\noindent The work is supported by the Office of Naval Research (ONR) under the Multidisciplinary University Research Initiatives (MURI) Grant Award N000142412545 as part of the METHODS (Machine Learning Enabled Two-Phase Flow Metrologies, Models, and Optimized Designs) program. The authors would like to acknowledge the computational resources and services at the Center for Computation and Visualization (CCV), Brown University.

\section*{Code and Data Availability}
\noindent All codes will be made publicly available at \url{https://github.com/SiaK4/PIKAN_Boiling.git} upon publication. The experimental data are available at \cite{DVN/LKAQFW_2026}.\\

\bibliography{references}

\end{document}